\documentclass{moriond}

\def\be{\begin{equation}}
\def\ee{\end{equation}}
\def\bea{\begin{eqnarray}}
\def\eea{\end{eqnarray}}

\usepackage{bm}
\usepackage{latexsym}

\newcommand{\lam}{\lambda}
\renewcommand{\P}{P}

\renewcommand{\div}{\Delta_\varepsilon}
\newcommand{\der}{\partial}
\renewcommand{\L}{\mathcal{L}}
\newcommand{\F} {\mathcal{F}}

\renewcommand{\P} {\mathcal{P}}
\newcommand{\U} {\textbf{U}}
\newcommand{\V} {\textbf{V}}

\newcommand{\bpi} {\bm{\pi}}
\newcommand{\btau} {\bm{\tau}}

\def\a{\alpha}
\def\b{\beta}
\def\g{\gamma}
\def\Tr{{\rm Tr}}

\def\beq{\begin{equation}}
\def\eeq{\end{equation}}

\def \bt#1 {\blue{\bm{#1}}}
\newcommand{\ba} {\begin{equation}\begin{aligned}}
\newcommand{\ea} {\end{aligned}\end{equation}}
\def\bea{\begin{eqnarray}}
\def\eea{\end{eqnarray}}

\newcommand{\blue}[1]{\color{blue} \textbf{#1} \color{black}}

\def\l{\left(}
\def\r{\right)}

\newcommand{\Photo}{}

\begin{document}
\vspace*{4cm}
\title{ONE LOOP EFFECTIVE NONLINEAR LAGRANGIAN \\ WITH A LIGHT H-BOSON}

\author{K. KANSHIN}
\address{Dipartimento di Fisica e Astronomia ``G.~Galilei'', Universit\`a di Padova and \\
INFN, Sezione di Padova, Via Marzolo~8, I-35131 Padua, Italy}


\maketitle\abstracts{
We study one loop structure of the scalar sector of non-linear electroweak chiral Lagrangian (EWChL) with a light (composite) H-boson up to four derivatives. First, we introduce relevant Lagrangian terms in general parametrization of would-be Goldstone modes, taking into account potential and  finite mass of the scalar. Then we compute 1-, 2-, 3- and 4-point functions and perform complete off-shell renormalization of the processes considered. On the way we found new divergencies involving also the H-boson which cannot be absorbed by the parameters of chiral invariant Lagrangian. We have demonstrated explicitly how these divergencies can be removed by field redefinition, and therefore proved that they are non-physical and give no contribution to the on-shell amplitudes. As a physical result renormalization group equations are derived to be used for future H-boson data analyses.}

\section{Introduction}
As far as discovery of H-boson has not been accompanied by appearance of further light states one seeks for the solution of so-called hierarchy problem, or in simple words why is it so light? Possible solution, H-boson as pseudo-goldstone boson was proposed long time ago~\cite{Kaplan:1983fs,Georgi:1984af} and nowadays received further development~\cite{Agashe:2004rs,Contino:2006qr}. EWChL is a model-independent way to describe the nonlinearly realized electroweak symmetry breaking enjoyed by those models. The present work is aimed to clarify the issues of consistency and completeness of the effective Lagrangian at loop level. It also prompts the expected size of coefficients of BSM effective operators. 
\section{The Lagrangian}
We consider the scalar sector of the EWChL invariant under global $SU(2)_L\times SU(2)_R$ chiral transformation. It has been previously studied~\cite{Grinstein:2007iv,Contino:2010mh,Azatov:2012bz,Azatov:2012qz,Brivio:2013pma} and fully derived in \cite{Buchalla:2013rka,Alonso:2012px}. In this work the notations of the last reference are adopted. The building blocks are the scalar field $h$ and $\V_\mu=(D_\mu \U) \U^\dag$ with $\U(\bpi)$ being the Goldstone bosons matrix  corresponding to the $SU(2)_{EW}\times~U(1)_Y\to U(1)_{em}$ symmetry breaking. Fields $\bpi$ denote triplet of "pions" -- longitudinal components of the gauge bosons.
 In the scalar sector the Lagrangian can be sorted according to the number of derivatives. In present work we go up to four derivatives:
\bea  \label{L}
\L&=&\L_0 + \L_2 + \L_4\,,
\\
- \L_0 &=& V(h)= 
  \mu_1^3 \,h + \frac{1}{2} m_h^{2} h^{2} + 
 \frac{\mu_3}{3!}  h^{3} +\frac{\lam}{4!} h^{4}\,,
\\
\L_{2} &=& 
\frac{1}{2} \der_\mu h \der^\mu h \ \F_H(h)
-\frac{v^2}{4} \Tr[\V_\mu  \V^\mu ] \ \F_C(h) \label{L2} \,, 
\\
\L_4&=&\sum_i c_i \P_i\,,
\eea
where functions $\F_i(h)=1+2a_i h/v+b_i h^2/v^2$ have to be treated as generic polynomials in h, coefficients $a_i, b_i$ encode deviation of H-boson from the doublet structure; $\mu_1$ is kept to cancel the tadpole divergence and set to zero after the renormalization. For explicit form of $\P_i$~-- terms with four derivatives and their expansions in terms of $\bpi$ we refer to the original paper \cite{Gavela:2014uta}. 

$\L_0,\L_2$ are used to derive both Feynman rules for the one loop calculation and correspondent counterterms, while $\L_4$ serves as a source of counterterms only.

Custodial symmetry breaking term with two derivatives has been omitted, since its coefficient is constrained to be small. Consequently for consistency at one loop level we do not need to take into account custodial breaking counterterms in $\L_4$ as we do not consider neither Yukawa terms nor gauge fields.   Therefore all the Lagrangian terms preserve custodial symmetry.

When relations between bare and renormalized parameters are set, counterterm Lagrangian is derived straightforwardly.
\subsection{General $\U$--matrix parametrization}
Requirement of $\U(\bpi)$ having proper transformation properties under chiral symmetry group does not fix completely the functional dependence on $\bpi$ field~\cite{Weinberg:1968de}. We consider expansion of $\U$ up to $\bpi^4$, since higher order terms do not contribute at one loop. In this case it can  be shown that the most general parametrization has the form:
\begin{equation}
  \U = 1-\frac{\bpi^2}{2v^2} - \l\eta+\frac{1}{8}\r \frac{\bpi^4}{v^4}
        + \frac{i (\bpi\btau)}{v} \l 1+ \eta \frac{\bpi^2}{v^2} \r + O(\bpi^5),
  \label{Uparam}
\end{equation}
where $\eta$ is unphysical "parametrization parameter". All the physical results will be independent of $\eta$, therefore general parametrization is a useful tool for the sanity check of the expressions obtained. 

Some particular choices of the parameter up to $O(\pi^5)$ correspond to the parametrizations widely used in literature: 
 $\eta=0$ gives the square root parametrization $\U =
\sqrt{1-\bpi^2/v^2}+i(\bpi\btau)/v$\,;
$\eta=-1/6$ corresponds to the exponential one:  
$\U =\exp(i\bpi\cdot\tau/v)$\,.

\section{Loops and divergencies}
We performed explicit computation of divergent parts and renormalization of all possible 1-, 2-, 3- and 4-point one loop Green functions involving $h$ and/or $\bpi$ off-shell external legs and found full agreement with previously known literature where only $\bpi$ legs and/or on-shell Green functions were considered \cite{Appelquist:1980ae,Delgado:2013hxa,Espriu:2013fia,Delgado:2014jda}. We adopted dimensional regularization and off-shell minimal subtraction scheme as renormalization procedure. 

By the off-shell renormalization we mean the matching of momenta structures with divergent coefficients generated by loops on the one hand with the momenta structures of the counterterms on the other. This procedure reveals the importance of some operators in $\L_4$ which are often disregarded in the literature. In our set up none of them can be disregarded or traded by equations of motion (EOM) unless full basis, including terms with fermions and gauge bosons is taken into account.

Finally, the divergent structures which cannot be matched with the any chiral invariant counterterms have been found, meaning this we call them non-invariant divergences (NIDs). However those divergencies do not vanish individually as the external legs are put on-shell, all the physical amplitudes, i.e. the combinations of all relevant Green functions giving divergent contribution to the process are NID free.  The problem of NID in nonlinear $\sigma$ model has been discussed long ago~\cite{Charap:1970xj,Kazakov:1976tj,Kazakov:1977mw,deWit:1979gw,Honerkamp:1971sh,Appelquist:1980ae}. Generalizing the approach of~\cite{Appelquist:1980ae} to the case of the light scalar in the spectrum we make pion field redefinition to remove  NIDs  from the final off-shell answer.
\subsection{Field redefinition}
It has been proved some time ago  that Lagrangians related by local field redefinitions, even including ones with space-time derivatives are physically equivalent~\cite{ostrogradsky1850memoire,GrosseKnetter:1993td,Scherer:1994wi,Arzt:1993gz}. In other words if redefinition $\bpi\to \bpi f(\bpi,h,\der_\mu)$ with $f(0)=1$ changes Lagrangian according to $\L\to\L+\delta\L$, then $\delta\L$ piece is unphyscial. Our goal is to find a proper $f$ to remove all NIDs. The minimal redefinition is:
\bea
&&\pi_i \to \pi_i \left( 1+ \frac{\a_1}{2v^4}\bpi\Box\bpi 
   + \frac{\a_2}{2v^4} \der_\mu\bpi\der^\mu\bpi+ \frac{\b}{2v^3}\Box h  \nonumber
   + \frac{\tilde{\g}_1}{2v^4}h\Box h+ \frac{\g_2}{2v^4}\der_\mu h\der^\mu h \right)+ \\ 
  & &\qquad + \frac{\alpha_3}{2v^4} \Box \pi_i (\bpi\bpi) 
   + \frac{\alpha_4}{2v^4}\der_\mu\pi_i (\bpi\der^\mu\bpi).    
\eea 
Treating $\delta\L$ as counterterm and matching it with NIDs we have obtained ($\Delta_\varepsilon$ is divergence):
\begin{equation}
\begin{array}{lll}
\begin{array}{l}
\a_1 = \left(9 \eta^2 + 5 \eta + \frac{3}{4}\right)\div, \\
\a_2 = \left[ 1+4\eta+\l \frac{1}{2}+\eta \r a_C^2 \right]  \div,\\
\a_3 = 2\eta^2 \div,\\
\a_4 = 2\eta \left(a_C^2-1\right)\div
\end{array}

&\quad&

\begin{array}{l}
   \beta = -\l\frac{3}{2}+5\eta\r a_C\div,\\
   \gamma_1 = \l\frac{3}{2}+5\eta\r \left(2 a_C^2-b_C\right)\div, \\
   \gamma_2 = \l\frac{3}{2}+5\eta\r \left(a_C^2-b_C\right)\div.\\
\end{array}

\end{array}
\label{eq:spa}
\end{equation}
Note that choice of $\eta=-3/10$ set all mixed $\bpi-h$ terms to zero. To our knowledge this does not correspond to any parametrization considered in the literature before. 

Thus we determined field redefinition which removes all the NIDs.
\subsection{Renormalization Group Equations}
After counterterms have been explicitly calculated it is straightforward to derive RGEs. For the resulting expressions we refer to the original paper~\cite{Gavela:2014uta}. It is worth mentioning that some of the terms in RGEs are weighted by large numerical factors, therefore even the couplings which are small at low energies can be enhanced to the large values by running to the high scales.
\section*{Acknowledgements}

The work is supported by an ESR contract of the European Union network FP7 ITN INVISIBLES (Marie Curie
Actions, PITN-GA-2011-289442), of MICINN, through the project
FPA2012-31880. 
I also would like to thank the organizers of Rencontres de Moriond meeting for the great organization of the event.

\end{document}